# Chromium isotopic homogeneity between the Moon, the Earth, and enstatite chondrites

**Bérengère Mougel[1], Frédéric Moynier[1,2], and Christa Göpel[1]**

[1]*Institut de Physique du Globe de Paris, Université Sorbonne Paris Cité, CNRS UMR 7154, Paris, France*

[2] *Institut Universitaire de France and Université Paris Diderot, Paris, France*

Corresponding: mougel@ipgp.fr

**ABSTRACT**

Among the elements exhibiting non-mass dependent isotopic variations in meteorites, chromium (Cr) has been central in arguing for an isotopic homogeneity between the Earth and the Moon, thus questioning physical models of Moon formation. However, the Cr isotopic composition of the Moon relies on two samples only, which define an average value that is slightly different from the terrestrial standard. Here, by determining the Cr isotopic composition of 17 lunar, 9 terrestrial and 5 enstatite chondrite samples, we re-assess the isotopic similarity between these different planetary bodies, and provide the first robust estimate for the Moon. In average, terrestrial and enstatite samples show similar $\varepsilon^{54}$Cr. On the other hand, lunar samples show variables excesses of $^{53}$Cr and $^{54}$Cr compared to terrestrial and enstatite chondrites samples with correlated $\varepsilon^{53}$Cr and $\varepsilon^{54}$Cr (per 10,000 deviation of the $^{53}$Cr/$^{52}$Cr and $^{54}$Cr/$^{52}$Cr ratios normalized to the $^{50}$Cr/$^{52}$Cr ratio from the NIST SRM 3112a Cr standard). Unlike previous suggestions, we show for the first time that cosmic irradiation can affect significantly the Cr isotopic composition of lunar materials. Moreover, we also suggest that rather than spallation reactions, neutron capture effects are the dominant process controlling the Cr isotope composition of lunar igneous rocks. This is supported by the correlation between $\varepsilon^{53}$Cr and $\varepsilon^{54}$Cr, and $^{150}$Sm/$^{152}$Sm ratios. After correction of these effects, the average $\varepsilon^{54}$Cr of the Moon is indistinguishable from the terrestrial and enstatite chondrite materials reinforcing the idea of an Earth-Moon-Enstatite chondrite system homogeneity. This is compatible with the most recent scenarios of Moon formation suggesting an efficient physical homogenization after a high-energy impact on a fast spinning Earth, and/or with an impactor originating from the same reservoir in the inner proto-planetary disk as the Earth and enstatite chondrites and having similar composition.

## 1. INTRODUCTION

The mechanism at the origin of the Earth-Moon system that is the most consistent with chemical and physical observations is through a giant impact(s) with the proto-Earth (Canup, 2004; Canup et al., 2012; Ćuk and Stewart, 2012; Ćuk et al., 2016). Canonical numerical simulations predicted that significant amount of the impactor should reside in the Moon, and thus that it could be detected in lunar samples. However, elements that display large nucleosynthetic anomalies between solar system materials (e.g., O, Ti, Cr) are isotopically similar in terrestrial and lunar samples (e.g. Qin et al., 2010; Zhang et al., 2012, Young et al., 2016), therefore questioning canonical models of Moon's formation suggesting that most of the Moon was composed of impactor materials (Canup, 2004). The first scenario that has been suggested to reconcile geochemical observations and canonical numerical simulations was the post-impact isotopic homogenization between the proto-Earth and the Moon through silicate vapour exchange (Pahlevan and Stevenson, 2007). However, while isotopic equilibrium can be achieved this way for volatile elements like O during the cooling timescale in the aftermath of the impact, it is almost impossible for the most refractory elements such as Ti (Tc=1582 K, with Tc the 50% condensation temperature), Fe (Tc=1334 K), Cr (Tc=1296 K) (e.g. Lodders, 2003; Zhang et al., 2012; Sossi and Moynier 2017). Based on the Ti isotopic similarity between the Earth and the Moon, it was therefore proposed that the Moon consists almost exclusively of terrestrial material (Zhang et al., 2012) unlike what is predicted by the canonical impact model. Alternatively, more recent studies also proposed that the Moon-forming impactor had an isotopic composition very close to that of the proto-Earth, and the only group of meteorites with a similar isotopic composition, the enstatite chondrites (e.g. Jacobsen et al., 2013; Dauphas et al., 2014a,b). In order to account for the isotopic similarities between the Earth-Moon system and enstatite chondrites (EC), it has been suggested that these materials initially came from the same uniform reservoir located in the inner part (≤1.5 AU) of the proto-planetary disk, and only lately diverged in their chemical evolution (e.g. Jacobsen et al., 2013; Dauphas et al., 2014a,b). This hypothesis would be consistent with both geochemical observations and state of the art of numerical simulations. However, a new set of numerical calculations that are based on a fast spinning Earth and a high energy impact (Ćuk and Stewart 2012; Lock et al., 2016; Ćuk et al., 2016) have opened up again the possibility of a high degree of post impact isotopic homogenisation between the Earth and the Moon, through the direct condensation of the Moon from an already well-mixed bulk silicate Earth vapour disk (Lock et al., 2016). In this model, this type of high energy impact leaves a system in a specific state where the Earth's vaporised mantle and the disk forms a homogenous medium from which the Moon grows and equilibrates. This scenario could also account for the isotopic similarity observed between the Earth and the Moon.

Recently, small mass-independent O isotopic offsets between the Earth and the Moon have been reported and interpreted as evidence for the detection of the impactor (Herwartz et al., 2014), however, these results have subsequently been questioned (Young et al., 2016).

Furthermore, although the Cr isotope compositions (expressed as per 10,000 deviation of the $^{53}Cr/^{52}Cr$ and $^{54}Cr/^{52}Cr$ ratios normalized from the NIST SRM 3112a standard) have been used as an argument in favor of the Earth-Moon system homogeneity, the limited dataset available for the Moon indicates a small isotopic difference between the two planetary objects (Qin et al., 2010; Warren, 2011). The Moon shows a slightly higher $\varepsilon^{54}Cr$ (i.e. sample 12061: $\varepsilon^{54}Cr = 0.22 \pm 0.10$, Qin et al., 2010) compared to the terrestrial standard, and enstatite chondrites ($\varepsilon^{54}Cr = 0.02 \pm 0.11$, 2sd, n=10, Trinquier et al., 2007; Qin et al., 2010). This could either be interpreted as an indicator for the nature of impactor, or it may be caused by the modification of the initial Cr isotopic composition through secondary processes. However, this average composition is based on the analysis of two lunar basalts only (Qin et al., 2010), and therefore it does not represent a robust estimate of the lunar Cr isotopic composition. Furthermore, the number of Cr isotopic measurements of terrestrial materials is relatively scarce (Trinquier et al., 2007; Qin et al., 2010), and all samples seem systematically slightly enriched in $^{54}Cr$ ($\varepsilon^{54}Cr = 0.00 \pm 0.08$ to $0.20 \pm 0.15$, n=8) compared to the terrestrial standard (NIST SRM 3112), as previously observed for Ni (e.g. Elliott and Steele, 2017) and Sr (Moynier et al., 2012). In order to evaluate if a possible isotopic difference between the Earth and the Moon does really exist, we investigate for the first time a much broader set of lunar samples (n=17) from various geological settings (mare basalts, ferroan anorthosites, Mg-suite, regoliths) and Apollo missions (12, 15, 16 and 17) together with nine terrestrial samples including basalts and peridotites and five enstatite chondrites.

## 2. SAMPLES AND METHODS

### 2.1 Analytical procedure

All chemical separations and isotopic measurements were performed at the Institut de Physique du Globe de Paris, France. Dozens to hundreds of mg of bulk rock samples were dissolved in an adjusted volume of concentrated HF and $HNO_3$ mixture (ratio 3:1) in Teflon bombs at 140°C for several days until complete dissolution, including multiple ultrasonication steps. No residual gels and/or refractory phases were observed during the final inspection of the solution. This dissolution procedure has previously been tested with success on pure chromite grains (Gopël et al., 2015). From these dissolutions, aliquots (equivalent to a mass ≤ 100 mg) were pipetted in order to target about 10 μg of Cr after separation on a single column. The chemical procedure adopted from Trinquier et al. (2008) includes 3 separation steps of Cr on cationic exchange resin AG50W-X8, and allows a ≥ 99% yield. $^{53}Cr/^{52}Cr$ and $^{54}Cr/^{52}Cr$ isotope ratios were measured by using a Fisher Scientific multi-collection (9 cups) Thermal-Ionization Mass-Spectrometry (TIMS) Triton. The details of the Cr isotopic measurements were published by Göpel et al. (2015). A purified Cr fraction, corresponding to 1-2 μg, was loaded in chloride form on degassed W filaments together with an Al-silicagel-$H_3BO_3$ emitter in order to facilitate Cr emission and stability. A typical measurement comprises 20 blocks of 10 cycles. Each single run (indicated as the letter n in Table 1) represents a combination of 3 successive multi-collection measurements with $^{53}Cr$, $^{52}Cr$ and $^{54}Cr$ isotopes in the center cup. The isotopic ratios obtained with each configuration represent independent measurements that can be averaged, compared to each other, and used to control the evolution of the instrument over time. $^{53}Cr/^{52}Cr$ and $^{54}Cr/^{52}Cr$ ratios of each beam configuration were normalized using an exponential law to $^{52}Cr/^{50}Cr = 19.28323$ (Shields et al., 1966). The $^{56}Fe$ intensity was monitored in order to control a possible isobaric interference of $^{54}Fe$ with $^{54}Cr$. All samples were loaded at least two times on filaments. Chemistry and measurements have been duplicated for lunar sample 15555, and give similar isotopic value (Table 1). Our data for enstatite chondrites and terrestrial rock standard BHVO-2 are consistent with earlier work (Trinquier et al., 2007; Qin et al., 2010) (Figure 1). Final Cr isotopic data are given in ε-units that represent the relative deviation in parts per 10,000 of

$^{53}Cr/^{52}Cr$ and $^{54}Cr/^{52}Cr$ ratios normalized to a terrestrial standard (NIST SRM 3112a Cr standard). All uncertainties are 2 standard errors (2 s.e.).

## 2.2 Samples overview

The chromium isotopic compositions of 17 lunar samples from Apollo 12, 14, 15, 16 and 17 missions (samples allocated by the CAPTEM committee) have been analyzed: 5 basalts (12002, 12005, 12040, 12063, 15555), 4 anorthosites (60015, 62255, 65315, 67955), one norite (77215), one dunite (72415), and 6 regolith samples (14163, 15041, 60501, 64501, 65701, 78221).

The basalts contain chromites mostly found as inclusions in pyroxenes and olivines, less frequently in the mesostasis (El Goresy et al., 1971). Samples 12005 and 12063 also show evidence for micrometeorite craters at their surface (Bloch et al., 1971). Anorthosites are cataclastic breccias but they have conserved pristine elemental compositions (Warren and Wasson, 1977). They exhibit systematically younger ($\leq$ 50 My) exposure ages than basalts (90 - 300 My). Although some of them show traces of micro-impacts, their meteoritic siderophile elements content is low. Sample 62255 is atypical because it contains a very large amount of impact melt ($\geq$ 35%; Ryder and Norman, 1980). The regolith represents admixtures of various components (fused material, plagioclase, mafic clasts, glass, highland and mare lithics) with different degrees of maturity; the average size of the grains ranges from 43$\mu$m (78220) up to 100$\mu$m (64501), and Is/FeO index (Is: relative concentration of nanophase metallic iron with ferromagnetic resonance) is between 55 (14163) and 110 (65701). Five enstatite chondrites including 3 EH (Qingzhen, Abee, Kota-Kota) and 2 EL (The Eagle, MAC88184) types were also studied. Finally, 9 terrestrial samples including basaltic references (BHVO-2, BE-N, NIST688), an East Pacific Rise basalt (10PUB22-07), and basalts from Cape Verde (CV-SN-98-19) and the Azores (KBD408729), as well as 2 peridotites from the Balmuccia massif (BM23, BM26) and one from the Kola peninsula (KOLA15-UB) were analyzed (see Rivalenti et al., 1995; Verhulst et al., 2000; Mourão et al., 2012; Mougel et al., 2014 for details on the sampled areas).

## 3. RESULTS

Chromium isotopic data are reported in Table 1. Our results for terrestrial samples and enstatite chondrites are illustrated in Figure 1. They are in good agreement with previous data from the literature, but show a slightly better precision. The $\varepsilon$ $^{54}$Cr values of terrestrial samples overlap those of EC samples, and range between -0.07 ± 0.07 and 0.19 ± 0.07. The $\varepsilon$ $^{53}$Cr values of EC are slightly higher $\varepsilon^{53}$Cr ($\varepsilon^{53}$Cr ≥ 0.10) than terrestrial compositions ($\varepsilon^{53}$Cr ≤ 0.10). $\varepsilon^{53}$Cr and $\varepsilon^{54}$Cr of lunar samples range from 0.01 ± 0.03 to 0.72 ± 0.04, and 0.06 ± 0.05 to 1.98 ± 0.09, respectively. These isotopic compositions strongly correlate for lunar samples ($R^2$=0.99) with a $\varepsilon^{54}$Cr/ $\varepsilon^{53}$Cr ratio of ~2.62 (Figure 2). The regoliths show systematically higher $\varepsilon^{54}$Cr than mare basalts, which exhibit higher $\varepsilon^{54}$Cr than noritic and anorthositic samples (except impact melt- rich 62255 sample). The dunite (72415) also shows a $\varepsilon^{54}$Cr value that falls in the range of mare basalts and anorthosites, and plots on the same correlation line as the rest of the samples in a $\varepsilon^{54}$Cr vs $\varepsilon^{53}$Cr diagram. Despite their different locations and modal composition variations, the regoliths form 3 pairs of similar Cr isotopic compositions that are consistent with maturity indices (i.e., grain size and Is/FeO). The distribution of the terrestrial samples is more scattered than lunar samples, however they appear to plot on the same trend as the lunar samples. In contrast to lunar samples, $\varepsilon^{53}$Cr and $\varepsilon^{54}$Cr values do not correlate in enstatite chondrites.

## 4. DISCUSSION

The homogeneous distribution of O and Ti isotopes in lunar samples (e.g. Wiechert et al., 2001; Zhang et al., 2012; Herwartz et al., 2014; Young et al., 2016) together with the correlation between $\varepsilon^{54}Cr$ and $\varepsilon^{53}Cr$ (Figure 2) suggests that the Cr isotopic variations observed in lunar rocks must be caused by secondary processes. $^{53}Cr$ excesses are due to the decay of $^{53}Mn$ ($T_{1/2}$=3.7 Myr; Honda and Imamura, 1971) whereas $^{54}Cr$ anomalies are controlled by the non-homogeneous distribution of different stellar components carrying various Cr isotopic compositions (Trinquier et al., 2007; Dauphas et al., 2010; Qin et al., 2011; Moynier et al. 2010), therefore they are not expected to be correlated. The correlation between $\varepsilon^{54}Cr$ and $\varepsilon^{53}Cr$ (Figure 2) in lunar samples may reflect the mixing between the proto-Earth and the giant impactor. However, the homogeneous composition of lunar rocks in respect to other isotopes displaying nucleosynthetic anomalies, as well as the fact that some lunar samples have $\varepsilon^{54}Cr$ and $\varepsilon^{53}Cr$ values higher than all other meteorite types (Göpel et al., 2015) implies again that Cr isotopes suffered from secondary processes. Late arrivals of meteoritic materials, solar wind implantation or galactic cosmic ray (GCR) irradiation are three possible candidates that might affect the Cr isotope abundances. The impact of these processes is discussed in the following paragraphs.

The presence of a continuous regolith and craters represent evidence for the erosion of the lunar surface by impacts (McKay et al., 1991). The arrival of meteoritic materials leads to the addition over time of non-lunar Cr to the Moon. As each type of meteorite is characterized by a distinct Cr isotopic composition, Cr isotopes can be used as a proxy for the composition of the infalling objects. This approach has been successfully applied to identify the nature of the impactors involved in the formation of many terrestrial impact structures (e.g., Koeberl et al., 2007; Kyte et al., 2011; Mougel et al., 2017; Magna et al. 2017). However, unlike terrestrial crustal rocks, mare basalts have on average chondrites-like abundances of Cr, implying that only very large meteoritic contributions (>20%) could be detected the lunar samples. Based on highly siderophile element data, it has been estimated that ~0.5-4% of meteoritic component was present in the lunar regolith (Korotev, 1987a; Norman et al., 2002; Puchtel et al., 2008). This amount is too low amount to generate detectable Cr isotopic variations. Moreover, because of 1) the variable Cr isotopic signature of meteorites (e.g. Göpel et al., 2015), 2) the multiple impactor components recorded in impact melt rocks (Puchtel et al., 2008; Fisher-Gödde and Becker, 2012; Liu et al., 2015) and 3) the fact that samples are from various areas and have different ages, no global correlation between samples in a $\varepsilon^{53}$Cr and $\varepsilon^{54}$Cr plot (Figure 2) would be expected. Therefore, the meteorite bombardment of the Moon cannot account for the variable Cr isotopic composition of lunar rocks.

Solar wind implantation represents the surface addition of low energy nuclei that are stopped inside the target material instead of inducing nuclear reactions within it. It is almost impossible to evaluate the composition of implanted Cr on the lunar surface because the Cr content and the isotopic composition of the solar wind are unknown. However, this effect must be negligible in bulk rock analysis because solar wind implantation process is unlikely to be effective deeper than the nano- to micro-meter scales as confirmed by noble gases implantation experiments (Grimberg et al., 2006; Bajo et al., 2015), and Cr leaching measurements on lunar soil grains (Kitts et al., 2003).

On the other hand, there is evidence for cosmogenic products accumulated in lunar rocks at depth greater than the centimeter scale (Bogard and Hirsch, 1975; Hidaka et al., 2000). This is due to nuclear reactions between high-energy particles from galactic cosmic rays (GCR) with exposed lunar materials. Long-term exposure to GCR can generate positive $\varepsilon^{53}$Cr and $\varepsilon^{54}$Cr shifts in planetary object (e.g., Shima and Honda, 1966; Trinquier et al, 2007; Qin et al., 2010; Leya et al., 2003). Cosmogenic addition of Cr isotopes can be produced through different reactions, which are mostly spallation on Fe, and spallation and neutron capture reactions on Cr itself. The spallation of Fe is usually considered as the dominant reaction for the production of cosmogenic Cr; it depends on the exposure age and the Fe/Cr ratio of the sample (Birck and Allègre, 1985b; Shima et Honda, 1966). Iron meteorites are the best examples for this effect. They have long exposure ages, very high Fe/Cr ratios, and exhibit large and correlated isotopic excesses in $^{54}Cr$ and $^{53}Cr$ produced with a $\varepsilon^{54}Cr$ / $\varepsilon^{53}Cr$ ratio of ~ 4, as described for Carbo meteorite pieces (Qin et al., 2010). These measurements were used to compare and confirm model predictions and estimations of GCR irradiation effects (Shima and Honda, 1966). For samples with low Fe/Cr ratios such as lunar samples, the Fe spallation effect becomes no longer significant, thus the effect of neutron captures and/or spallation reactions on Cr may be observed (Leya et al., 2003). However, no large Cr isotopic excesses were previously reported in bulk lunar rocks before our study, therefore it was concluded that none of these cosmogenic reactions significantly affect Cr isotopes in lunar material (Leya et al., 2003; Qin et al., 2010). The first evidence for the existence of Cr isotopic anomalies in lunar material arose from the study of Cr-depleted minerals (i.e. plagioclase) from Apollo 16 regoliths (62281 and 60601). Leaching experiments on individual plagioclase grains have shown surface-correlated $\varepsilon^{53}Cr$ and $\varepsilon^{54}Cr$ highly positive anomalies that have been attributed to solar wind implantation effect (Kitts et al., 2003). However, this hypothesis remains to be verified by further investigations, and in any case, it is negligible at the bulk rock scale, thus leaving cosmogenic production of Cr the most probable mechanism that can account for the shifts of Cr isotopes towards positive values, as observed in bulk lunar samples measured in this study. Two further observations also support this idea. First, the samples that have the highest exposure ages (Meyer, 2004) exhibit in general higher $\varepsilon^{53}Cr$ and $\varepsilon^{54}Cr$ than the samples with the lowest exposure ages (except 62255). Second, based on the separate measurements of two distinct fractions of the same sample (12002), we show that at the centimeter scale Cr isotope ratios may vary within an igneous rock (Table 1), which is together with the first observation consistent with irradiation effects. Because the exposure ages of lunar rocks are poorly constrained, or undetermined, it is difficult to apply traditional methods of correction of Fe spallation effect (Birck and Allègre, 1985b; Trinquier et al., 2007; Qin et al., 2010) in order to calculate the pre-irradiation Cr isotopic composition of the Moon. Moreover, the Fe/Cr ratios of the lunar samples measured here are relatively low (average of ~ 45) and are not positively correlated with Cr isotopic compositions implying that the cosmogenic addition of Cr through Fe spallation effect may actually be negligible. Based on the GCR model of Leya et al. (2003), which expresses $\varepsilon_{GCR}^{53}Cr$ (shift on $\varepsilon^{53}Cr$ induced by GCR) as a function of $\varepsilon_{GCR}^{54}Cr$ (shift on $\varepsilon^{54}Cr$ induced by GCR) for exposure ages between 100 and 500 Myr, the slope of the correlation for a Fe/Cr ratio of 45 can be estimated and compared to our results. This GCR-model prediction is represented in Figure 2, and is inconsistent with the Cr isotopic data obtained for lunar samples, thus suggesting that the cosmogenic effects recorded at the surface of the Moon are not well estimated for Cr. These reactions depend on various parameters including nuclide cross-sections, the energy of the incident particles, and the energy profile of the target material. The uncertainties on some of these parameters remain high. For example, cross-sections used in the model of Leya et al. (2003) are theoretical values that are expected to be accurate to within a factor of 2 only. Whereas these uncertainties are on not significantly changing the isotopic ratios for medium to high-energy spallation reactions, they may have more implications on the estimate of lower energy neutron capture effect.

Isotopic excesses in $^{150}Sm/^{152}Sm$ and $^{158}Gd/^{160}Gd$ ratios are other good proxies for cosmogenic effects induced from GCR-interaction with planetary materials. They are tracers of neutron capture reactions on $^{149}Sm$ and $^{157}Gd$, which have extremely large capture cross-sections (Eugster et al., 1970; Lugmair and Marti, 1971; Russ, 1972). Since the $^{158}Gd/^{160}Gd$ compositions corresponding to our samples dataset have not been measured yet, therefore only $^{150}$Sm/$^{152}$Sm isotopic compositions are discussed here. Cosmic rays are mostly made of high-energy protons and alpha particles producing secondary neutrons of low energies that can be captured by some isotopes. These reactions have always been thought to have only a minor impact on $^{53}$Cr/$^{52}$Cr and $^{54}$Cr/$^{52}$Cr ratios compared to spallation (Leya et al., 2003). However, when plotting the Cr isotopic compositions of the lunar bedrocks measured here versus their corresponding Sm isotopic compositions (Nyquist et al., 1995; Sands et al., 2001; Boyet and Carlson, 2007; McLeod et al., 2014; Carlson et al., 2014; Boyet et al., 2015; Albalat et al., 2015) it is clear that both systems are directly coupled (Figure 3). Both $\varepsilon^{53}$Cr and $\varepsilon^{54}$Cr values are linearly correlated with $^{150}$Sm/$^{152}$Sm ratios, thus suggesting that neutron capture process for the lunar materials plays a more important role than previously thought (Leya et al., 2003), and observed in meteorites (Qin et al., 2010) in controlling Cr isotopic compositions. The very weak slope of the Sm-Cr correlation (Fig. 3) also seems at first order consistent with the difference in neutron capture cross-section values that are orders of magnitude higher for $^{149}$Sm than $^{52}$Cr and $^{53}$Cr isotopes. Therefore, the discrepancy between theoretical models and measurements presented most likely comes from the underestimation of the impact of neutron capture effects on Cr isotopes for low Fe/Cr materials.

The lunar regoliths do not fall on this Sm-Cr trend defined by the rest of the samples, and do not correlate together (Figure 3). However, they fall on the same correlation as the rest of the samples in a $\varepsilon^{53}$Cr - $\varepsilon^{54}$Cr diagram (Figure 2). Depending on the type of process involved (e.g. spallation, neutron capture, solar wind implantation, mixing with meteoritic material), different excess production $\varepsilon^{54}$Cr/$\varepsilon^{53}$Cr ratios are expected because these processes are not controlled by the same parameters. The fact that we observe a single correlation in lunar samples between $\varepsilon^{53}$Cr vs. $\varepsilon^{54}$Cr advocates for a single dominant process affecting Cr isotopes, unless the contributions of all the processes are identical in all the samples, which is very unlikely. The decoupling between Sm and Cr isotope systems in regolith samples can be explained by the combination of two factors: 1) The different nature and behavior of the Sm and Cr carrier phases; and 2) The complex formation and re-working of the regolith. Lunar regoliths can be considered as admixtures of polygenic and unconsolidated micrometric fragments, of various degrees of maturity and homogenization. They are likely more sensitive to isotopic variations at a small scale than larger bedrock fragments, because their bulk composition also depends on the distribution and amount of isotope-carrier grains in the fraction that is analyzed. Lunar melts rocks are enriched in Cr (1000 – 6000 ppm) but the carrier phases of Cr are restricted to Cr-spinel and pyroxenes (Bonnand et al., 2016). Pyroxene is much more abundant but dramatically less enriched in Cr than chromite, which essentially controls the Cr budget of a lunar rock. The Sm concentration is relatively constant (2 – 20 ppm) in lunar rocks. It is an incompatible element that is commonly enriched in the mesostasis of lunar melts, and in late-stage crystallization phases such as phosphates (e.g. Shearer et al., 2015). Because chromite is more resistant than phosphate and mesostasis, Sm is expected to be more easily re-mobilized during regolith maturation than Cr, which again supports a Sm - Cr isotopic decoupling in regoliths.

By extrapolation of the linear regressions (Figure 3) to the "non-cosmogenic" value of $^{150}Sm/^{152}Sm$ (0.2760), we can deduce corrected Cr isotopic compositions for the Moon of $\varepsilon^{53}Cr = 0.03 \pm 0.04$ and $\varepsilon^{54}Cr = 0.09 \pm 0.08$. These values are similar to the average composition of the lunar samples that show terrestrial Sm isotopic compositions ($\varepsilon^{53}Cr = 0.02 \pm 0.02$ and $\varepsilon^{54}Cr = 0.08 \pm 0.06$). Based on this average value, we show that the $\varepsilon^{54}Cr$ value for the Moon is identical within error to the one of the Earth ($\varepsilon^{54}Cr = 0.10 \pm 0.13$) and to that of enstatite chondrite ($\varepsilon^{54}Cr = 0.02 \pm 0.11$). Our results therefore corroborate the idea that they are all originated in the same isotopic reservoir in the protoplanetary disk (Jacobsen et al., 2013; Dauphas et al., 2014a,b) that diverged through early Mn/Cr fractionation events, and explain the slight difference in $\varepsilon^{53}$Cr between EC and the Earth/Moon. Some cosmogenic effects on terrestrial rocks may also be responsible for the small scatter of the $\varepsilon^{53}$Cr and $\varepsilon^{54}$Cr observed in terrestrial rocks (Figure 1 and Figure 2) although we cannot truly confirm it. This is supported by the fact that they plot on the same correlation as lunar samples, and are systematically shifted toward positive values, indicating that the same process most likely affects them, in contrast to EC. However, the Sm cosmogenic variability for a wide range of terrestrial material has never been really investigated. If it turns out that similar cosmogenic effects affect the terrestrial Sm and Cr isotope compositions, the pre-irradiation $\varepsilon^{54}$Cr of the Moon and Earth may lie even closer to the composition of enstatite chondrites. Finally, the Cr isotopic similarity between the Earth, the Moon and the enstatite chondrite parent bodies can now be confidently added to the list of elements showing isotopic similarity between these planetary bodies (e.g, Ti: Zhang et al., 2012; O: Young et al., 2016). The isotopic similarity between the Earth and the enstatite chondrites had led to a series of models suggesting that the Earth accreted principally from enstatite meteorite like materials (e.g. Javoy, 1995; Javoy et al., 2010) or to a mixing between different types of chondrites (e.g. Lodders, 2000; Warren, 2011, Fitoussi and Bourdon, 2012; Dauphas, 2017). For example, Lodders (2000) proposed a mixing of 70% EH, 21% H, 5% CV and 4% CI to account for the composition of the Earth. Also considering that the bulk silicate Earth's Cr budget has recorded the last 85% of the total amount of material accreted by the Earth (Dauphas, 2017), this type of mixing would produce an average terrestrial $\varepsilon^{54}$Cr that would be in agreement with the average composition we report in this study ($\varepsilon^{54}$Cr = 0.10 ± 0.13). The Cr (this study) and Ti (Zhang et al., 2012) isotopic similarity between the Moon and the Earth fit also well with the recent models of the high-energy impacts predicting isotopic homogeneity between the Earth and the Moon (Lock et al., 2016; Ćuk et al., 2016).

## 5. CONCLUSIONS

We found that the Cr isotopic composition of most lunar rocks is not primary, but that it is affected by galactic cosmic ray irradiation that shifts the isotopic compositions towards higher $\varepsilon^{53}$Cr and $\varepsilon^{54}$Cr values. Lunar samples plot along a line in a $\varepsilon^{53}$Cr- $\varepsilon^{54}$Cr diagram with a slope of ~2.62, which is at odd with previous lunar GCR model predictions. It suggests that these effects have to be re-visited for Cr for the Moon and possibly other planetary materials. Furthermore, unlike in iron meteorites and other long time exposure chondrites, which can be affected by important Fe spallation effect, neutron capture appears to be the process that controls the Cr isotopic deviation of the lunar surface, based on the correlation of Cr and Sm isotopic compositions. After correction of the cosmogenic effects, we can now confidently state that the Moon $\varepsilon^{54}$Cr value (0.09 ± 0.08) is indistinguishable from the Earth (0.10 ± 0.13), and within error strictly similar to enstatite chondrites (0.02 ± 0.11). This further emphasizes either that the impactor at the origin of Moon had as a composition close to that of the Earth and enstatite chondrites, and thus may come from the same inner-disk reservoir, and/or that a high-energy impact followed by subsequent efficient physical mixing as suggested in the most recent numerical simulations is at the origin of the Moon.

**ACKNOWLEDGEMENTS**

We thank the ERC under the European Community's H2020 framework program/ERC grant agreement # 637503 (Pristine) and the ANR for a chaire d'Excellence Sorbonne Paris Cité (IDEX13C445) and for the UnivEarthS Labex program (ANR-10-LABX-0023 and ANR-11-IDEX-0005-02). Parts of this work were supported by IPGP multidisciplinary program PARI, and by Region île-de-France SESAME Grant no. 12015908. Editor William B. McKinnon and the two anonymous reviewers are thanked for the time spent on the manuscript, and their interesting and helpful comments.

**TABLE AND FIGURE CAPTIONS**

Table 1 – Chromium isotope compositions ($\varepsilon^{54}$Cr and $\varepsilon^{53}$Cr) of bulk lunar, terrestrial, and enstatite chondrite samples. The uncertainties of individual Cr isotopic analyses are 2 standard errors (n measurements). The letter n indicates the number of TIMS measurements of the sample. One measurement comprises 20 blocks of 10 cycles measured in 3-lines cup configuration. The notation $\varepsilon^{53}$Cr and $\varepsilon^{54}$Cr represents the relative deviation in parts per 10,000 of $^{53}$Cr/$^{52}$Cr and $^{54}$Cr/$^{52}$Cr ratios (normalized to $^{50}$Cr/$^{52}$Cr) from the NIST SRM 3112a Cr standard. * $^{150}$Sm/$^{152}$Sm (normalized to $^{147}$Sm/$^{152}$Sm) from data compilation from Nyquist et al., 1995; Boyet and Carlson (2007); McLeod et al. (2014); Carlson et al. (2014); Boyet et al. (2015). **Cr isotopic data from Qin et al., 2010.

Figure 1 – $\varepsilon^{53}$Cr versus $\varepsilon^{54}$Cr plot. Comparison of terrestrial (orange circles) and enstatite chondrite (purple circles) compared to data from Trinquier et al., (2007); and Qin et al., (2010) represented by squares (yellow and pink for terrestrial and enstatite chondrites respectively).

Figure 2 – Chromium isotopic compositions ($\varepsilon^{54}$Cr and $\varepsilon^{53}$Cr) of individual lunar samples (blue circles) represented together with the average compositions of terrestrial (orange square) and enstatite chondrite (purple square) data from this study (big circles) and literature (small

circles; Trinquier et al., 2007; Qin et al., 2010). The black line represents the linear regression related to lunar measurements. The dashed black line represents the predicted Cr isotopic composition for lunar material with cosmic-ray exposure ages of 100 to 500 Myr, and Fe/Cr = 45 (Leya et al., 2003).

Figure 3 – Correction of cosmogenic Cr isotopic effects in lunar samples (blue squares). $^{150}$Sm/$^{152}$Sm (normalized to $^{147}$Sm/$^{152}$Sm) data in Table 1 versus A) $\varepsilon^{53}$Cr, and B) $\varepsilon^{54}$Cr. Error bars for the Sm isotopic ratios are smaller than the size of the symbols. The yellow lines represent the non-cosmogenic Sm isotopic value (0.2660). The black lines are the linear regressions for lunar bedrock samples (62255 excluded, dashed circle). The lunar regolith samples are represented as triangles. The dark blue triangles symbolize the corrected Cr isotopic values for the Moon ($\varepsilon^{53}$Cr = 0.03 ± 0.04 and $\varepsilon^{54}$Cr = 0.09 ± 0.08).

TABLE 1

| Sample | Type | ε$^{53}$Cr | 2 s.e | ε$^{54}$Cr | 2 s.e | n | $^{150}$Sm/$^{152}$Sm * |
|---|---|---|---|---|---|---|---|
| *Lunar rocks* | | | | | | | |
| 12063 | basalt | 0.08 | 0.04 | 0.22 | 0.06 | 10 | n.a |
| 15555 | basalt | 0.07 | 0.03 | 0.20 | 0.05 | 14 | 0.27626 |
| 15555 dup | basalt | 0.07 | 0.07 | 0.22 | 0.14 | 5 | 0.27626 |
| 12016 | basalt | 0.1 | 0.06 | 0.22 | 0.1 | ** | 0.27636 |
| 70017 | basalt | 0.03 | 0.06 | 0.08 | 0.12 | ** | 0.27608 |
| 12002, 605 | basalt | 0.02 | 0.03 | 0.05 | 0.05 | 4 | n.a |
| 12002, 607 | basalt | 0.10 | 0.03 | 0.2 | 0.07 | 4 | 0.27645 |
| 12005 | basalt | 0.10 | 0.03 | 0.28 | 0.08 | 5 | 0.27722 |
| 12040 | basalt | 0.16 | 0.03 | 0.45 | 0.08 | 4 | 0.27786 |
| 67955 | anorthosite | 0.03 | 0.05 | 0.07 | 0.09 | 10 | 0.27602 |
| 60015 | anorthosite | 0.05 | 0.02 | 0.08 | 0.06 | 8 | n.a |
| 65315 | anorthosite | 0.02 | 0.02 | 0.12 | 0.04 | 8 | 0.27599 |
| 62255 | anorthosite | 0.13 | 0.02 | 0.35 | 0.05 | 9 | 0.27596 |
| 77215 | norite | 0.01 | 0.03 | 0.06 | 0.05 | 12 | 0.27606 |
| 72415 | dunite | 0.10 | 0.04 | 0.31 | 0.07 | 4 | n.a |
| 78221 | soil | 0.28 | 0.04 | 0.90 | 0.09 | 3 | n.a |
| 15041 | soil | 0.33 | 0.02 | 0.98 | 0.05 | 4 | n.a |
| 65701 | soil | 0.72 | 0.04 | 1.98 | 0.09 | 4 | n.a |
| 14163 | soil | 0.57 | 0.03 | 1.49 | 0.08 | 4 | 0.27745 |
| 60501 | soil | 0.72 | 0.04 | 1.82 | 0.1 | 4 | 0.27716 |
| 64501 | soil | 0.59 | 0.03 | 1.48 | 0.08 | 4 | 0.27915 |
| *Enstatite chondrites* | | | | | | | |
| The Eagle | EL6 | 0.14 | 0.05 | -0.07 | 0.07 | 5 | n.a |
| MAC88184 | EL3 | 0.20 | 0.03 | 0.11 | 0.07 | 4 | n.a |
| Abee | EH4 | 0.05 | 0.04 | -0.02 | 0.08 | 4 | n.a |
| Kota-Kota | EH3 | 0.11 | 0.04 | 0.00 | 0.08 | 6 | n.a |
| QingZhen | EH3 | 0.12 | 0.04 | 0.00 | 0.05 | 5 | n.a |
| *Terrestrial rocks* | | | | | | | |
| 10PUB22-07 | basalt | 0.08 | 0.05 | 0.19 | 0.07 | 8 | n.a |
| KBD408729 | basalt | 0.07 | 0.03 | 0.14 | 0.07 | 4 | n.a |
| NIST688 | basalt | 0.06 | 0.04 | 0.11 | 0.07 | 5 | n.a |
| CV-SN-98-19 | basalt | 0.01 | 0.05 | 0.12 | 0.11 | 4 | n.a |
| BE-N | basalt | 0.12 | 0.03 | 0.17 | 0.07 | 3 | n.a |
| BHVO-2 | basalt | 0.02 | 0.03 | 0.07 | 0.07 | 8 | n.a |
| KOLA15-UB | peridotite | 0.00 | 0.03 | 0.05 | 0.06 | 8 | n.a |
| BM31 | peridotite | 0.07 | 0.04 | 0.13 | 0.08 | 6 | n.a |
| BM23 | peridotite | -0.01 | 0.05 | 0.03 | 0.08 | 4 | n.a |

FIGURE 1

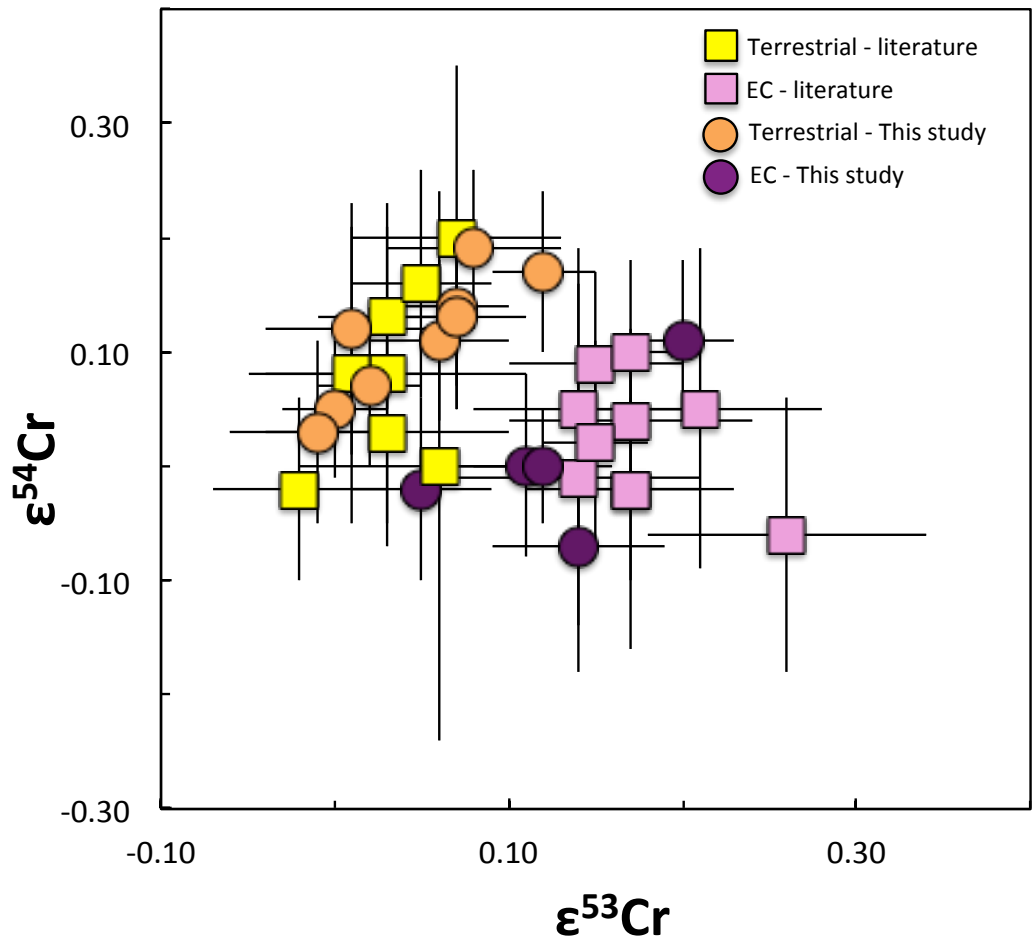

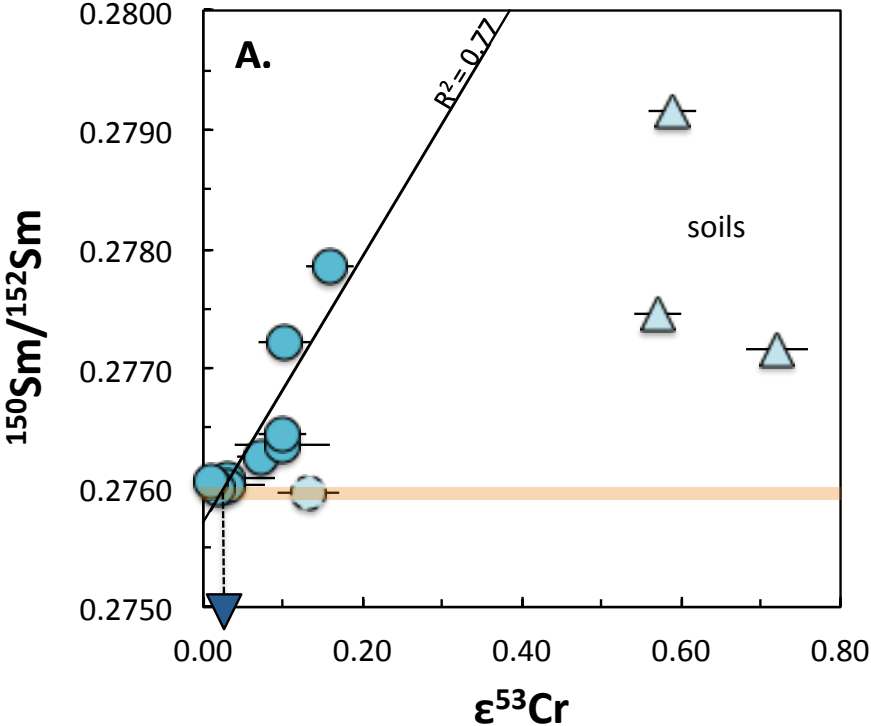

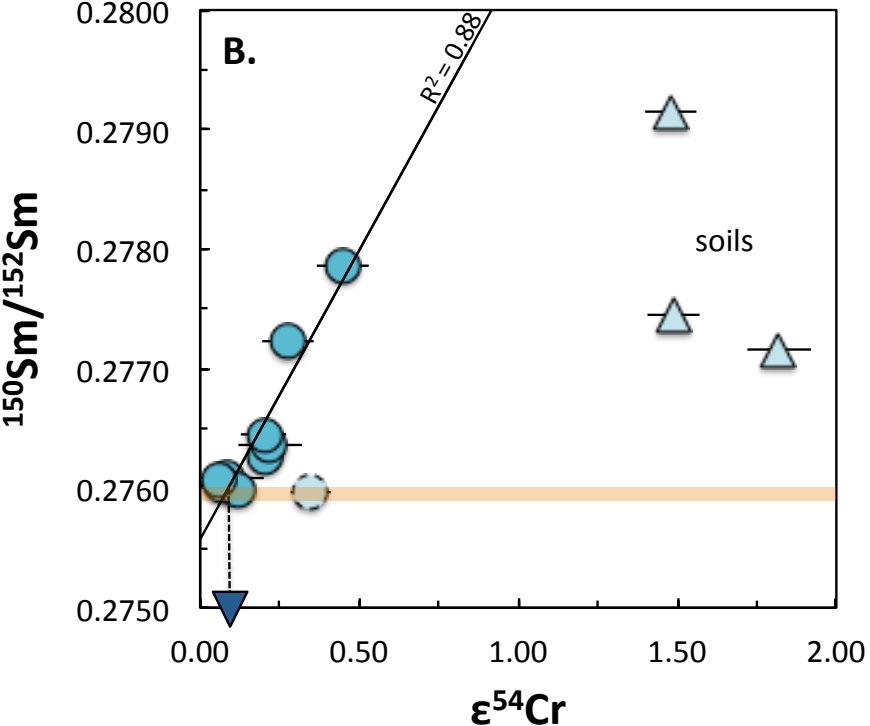